\documentclass[aps,prd,reprint,longbibliography,nofootinbib,superscriptaddress]{revtex4-2}

\usepackage{natbib}
\usepackage{slashed}
\usepackage{verbatim}
\usepackage[T1]{fontenc}
\usepackage{mathbbol}
\usepackage[dvipsnames]{xcolor}
\usepackage{orcidlink}
\usepackage[english]{babel}
\usepackage{lipsum}
\usepackage{array}
\usepackage{comment}
\usepackage{dcolumn}
\usepackage[normalem]{ulem}
\usepackage{url}
\usepackage{tensor}
\usepackage{comment}
\usepackage{graphicx,color,overpic,mathtools}
\usepackage{amsthm,amsmath,amssymb,mathrsfs}
\usepackage{soul}
\usepackage{microtype}
\usepackage{braket,bm,bbm,setspace}
\usepackage{cancel}
\usepackage{float}
\usepackage{xargs}
\usepackage{physics}
\definecolor{teal}{RGB}{0, 128, 128}
\definecolor{myred}{RGB}{179, 27, 27}
\usepackage{hyperref}
\hypersetup{
	colorlinks=true,
	linkcolor=myred, 
	citecolor=myred, 
	urlcolor=myred  
}

\newcommand{\sig}{\varsigma}

\newcommand{\Rone}{R_1^{(\sig)}}
\newcommand{\Kt}{\widetilde K}

\begin{document}
	
	\title{The quasinormal modes of Kerr–Newman black holes:\\ a separated approach}
	
	%
	\author{Kyriakos Destounis}
	\affiliation{CENTRA, Departamento de Física, Instituto Superior Técnico – IST, Universidade de Lisboa – UL, Avenida Rovisco Pais 1, 1049-001 Lisboa, Portugal}
	\affiliation{School of Applied Mathematical and Physical Sciences – SEMFE, Physics Department, National Technical University of Athens -- Polytechnic Campus, Zografou 15772, Athens, Greece}
	\author{Vitor Cardoso}
	\affiliation{CENTRA, Departamento de Física, Instituto Superior Técnico – IST, Universidade de Lisboa – UL, Avenida Rovisco Pais 1, 1049-001 Lisboa, Portugal}
	\affiliation{Center of Gravity, Niels Bohr Institute, Blegdamsvej 17, 2100 Copenhagen, Denmark}
	\author{Peter Hintz}
	\affiliation{Department of Mathematics, Pennsylvania State University,
		54 McAllister St, State College, PA 16801, United States}
	\begin{abstract}
		We compute Kerr–Newman quasinormal modes using a recently found exact separation of the gravitoelectric perturbation equations. For each spin label $\sig=\pm1$, a matrix angular ordinary differential equation and a scalar second-order radial ordinary differential equation define a joint spectral problem for the complex frequency $\omega$ and separation constant $\lambda$. We solve this problem by imposing regularity at the angular poles, ingoing behavior at the event horizon, and outgoing behavior at infinity. We benchmark the numerical code built to solve the separated equations, and the resulting frequencies, against recently published spectral calculations of the coupled system. Comparisons with those results yield absolute differences of order $10^{-10}$ to $10^{-5}$, depending on the BH parameters and overtone number. We also find numerical hints of spin-label isospectrality in both the electromagnetic-led and gravitational-led sectors of fundamental quasinormal modes, even at near-extremal KN configurations. These results provide numerical evidence that the separated equations reproduce Kerr–Newman quasinormal frequencies and offer a practical route to exploring the ringdown of rotating, charged black holes.
	\end{abstract}
	
	\maketitle
	
	\noindent {\bf \emph{Introduction.}} 
	Black-hole (BH) quasinormal modes (QNMs) describe the characteristic damped oscillations of a perturbed spacetime \cite{Kokkotas:1999bd,Berti:2009kk}. Their complex frequencies encode both the oscillation rate and the decay timescale, linking the dynamics of perturbations to the properties of the underlying BH. They play a central role in the study of BH mode stability and in the interpretation of gravitational-wave ringdown signals~\cite{Berti:2025hly}. Determining QNM spectra is, therefore, an important step toward understanding how BHs respond to perturbations and approach equilibrium.
	
	The Kerr–Newman (KN) BH solution provides the setting for studying the combined effects of rotation and electric charge in asymptotically flat Einstein–Maxwell theory \cite{Newman:1965my}. Its perturbations involve interacting gravitational and electromagnetic degrees of freedom, whose coupling makes the spectral problem substantially more challenging than in simpler BH backgrounds. Although astrophysical BHs are generally expected to carry negligible electric charge, the KN family remains a fundamental theoretical laboratory for exploring radiative dynamics in electrovacuum general relativity (GR). A tractable description of KN QNMs can clarify how rotation, charge, and field coupling shape the spectrum. Previous approaches to KN QNMs include the Dudley--Finley approximation \cite{Berti:2005eb,Saha:2025nsg,Silva:2025khf}, which decouples the coupled gravitoelectric system of equations, perturbative approximations in the slow-rotation limit \cite{Pani:2013ija,Pani:2013wsa,Blazquez-Salcedo:2022eik} and small-charge limit \cite{Mark:2014aja}, and fully numerical calculations, either via time-domain numerical methods~\cite{Zilhao:2014wqa} or through spectral methods that solve the coupled set of perturbation equations~\cite{Dias:2015wqa,Dias:2022oqm,Pombo:2026xsj}.
	
	Recently, an exact method was found that separates the perturbation equations, without approximations \cite{Hintz:2026jiw}. This milestone offers a promising route to the KN QNM problem. Separation reduces the coupled dependence on the radial and angular coordinates to a joint eigenvalue problem for ordinary differential equations, while retaining the interaction between gravitational and electromagnetic perturbations. The angular and radial problems remain linked through the complex frequency and separation constant, which must simultaneously satisfy angular regularity and radiative boundary conditions. This formulation makes established numerical methods for ordinary differential equations available and allows the angular and radial resolutions to be controlled independently. It also facilitates systematic exploration of the BH parameter space and tracking of individual modes as the rotation and charge vary. Beyond numerical computation, the separated equations provide a starting point for analytical approximations and investigations of limiting regimes and possible QNM symmetries.
	
	The aim of this work is to investigate the applicability of the novel formulation \cite{Hintz:2026jiw} to KN QNMs. We compute the complex frequencies and their associated angular eigenvalues, assess numerical convergence, and compare the results with independent calculations of the coupled gravitoelectric perturbation system \cite{Dias:2015wqa,Pombo:2026xsj,pombo_2026_21359120}. These comparisons test whether the separated equations recover the physical QNM frequencies across the configurations studied and establish their practical value for ringdown calculations.

	\noindent {\bf \emph{Kerr--Newman spacetime.}} 
	We work in geometrized units, $G=c=1$, and consider a KN BH in the Boyer-Lindquist coordinates $(t,r,\theta,\phi)$, with metric signature $(+,-,-,-)$, i.e.,
	\begin{equation}
		\begin{aligned}
			ds^2 &=
			\frac{\Delta-a^2\sin^2\theta}{\Sigma}\,dt^2
			+\frac{2a(2Mr-Q^2)\sin^2\theta}{\Sigma}\,dt\,d\phi
			\\
			&\!-\frac{\Sigma}{\Delta}\,dr^2
			-\Sigma\,d\theta^2-\frac{(r^2+a^2)^2{-}a^2\Delta\sin^2\theta}{\Sigma}
			\sin^2\theta\,d\phi^2,
		\end{aligned}
		\label{eq:kn-metric}
	\end{equation}
	where
	\begin{equation}
		\Delta(r)=r^2-2Mr+a^2+Q^2,
		\quad
		\Sigma(r,\theta)=r^2+a^2\cos^2\theta.
		\label{eq:delta-sigma}
	\end{equation}
	Here, $M>0$ is the mass, $Q>0$ is the electric charge, and $a=J/M$ is the angular momentum per unit mass. Subextremal KN BHs possess two horizons
	\begin{equation}
		r_\pm=M\pm\sqrt{M^2-a^2-Q^2},
	\end{equation}
	where $r=r_+$ and $r=r_-$ are the event and Cauchy horizons, respectively. The time and azimuthal dependence of modes is $e^{-i\omega t+im\phi}$, with $\omega\in\mathbb C$ and $m\in\mathbb Z$. A damped mode has, therefore, $\operatorname{Im}\omega<0$. Our numerical treatment of the equations of Ref. \cite{Hintz:2026jiw} scales out the mass and works with
	\begin{equation}
		\hat r=r/M,\quad \hat a=a/M,\quad \hat Q=Q/M,
		\quad \hat\omega=M\omega.
	\end{equation}
	
	\noindent {\bf\emph{Angular eigenvalue problem.}} 
	In Appendix~\ref{app:angular-basis}, we show that the angular eigenvalue problem in Ref.~\cite{Hintz:2026jiw} is equivalent to
	\begin{equation}
		\begin{pmatrix}
			T & i(\partial_\theta+W)\\
			i(\partial_\theta-W)&-T
		\end{pmatrix}
		\begin{pmatrix}\chi_1\\\chi_2\end{pmatrix}
		=\lambda\begin{pmatrix}\chi_1\\\chi_2\end{pmatrix},
		\label{eq:angular-two}
	\end{equation}
	for complex-valued functions $\chi_1,\chi_2$ of $\theta\in(0,\pi)$, where
	\begin{align}
		W(\theta)&=\frac32\cot\theta+
		\sig\left[\frac{m}{\sin\theta}
		-\frac{a\omega(M^2+Q^2)}{M^2-Q^2}\sin\theta\right],\\
		T(\theta)&=-\frac{3M}{2Q}
		-\sig\frac{2a\omega MQ}{M^2-Q^2}\cos\theta.
		\label{eq:angular-coefficients}
	\end{align}
	The label $\sig$ distinguishes the positive- and negative-spin descriptions used in Ref.~\cite{Hintz:2026jiw}; it does not label gravitational-led versus electromagnetic-led modes. The operator in~\eqref{eq:angular-two} is self-adjoint on its appropriate regular domain for real $\omega$, but generally not for complex $\omega$. For the QNM problem, we allow the frequency $\omega$ and separation constant $\lambda$ to be complex.
	
	At $a=0$, the two branches for $\ell\ge2$ and
	$|m|\le\ell$ satisfy
	\begin{equation}
		\lambda_{\ell,b}^{(0)}
		=b\sqrt{(\ell-1)(\ell+2)+\frac{9M^2}{4Q^2}},
		\qquad b=\pm1.
		\label{eq:spherical-lambda}
	\end{equation}
	Thus $\lambda$ is not the scalar-Laplacian eigenvalue $\ell(\ell+1)$ when $a=0$. The label $b$ specifies a spherical angular branch, followed continuously as parameters change; it is distinct from $\sig$.

	\noindent {\bf\emph{Separated radial equations.}} 
	We put
	\begin{equation}
		K=-(r^2+a^2)\omega+am, \ 
		\xi_0=\frac{2\omega Q^2}{M^2-Q^2}, \ 
		\Kt=K-\xi_0\Delta,
		\label{eq:radial-defs}
	\end{equation}
	and 
	\begin{equation}
		\Rone=-\sig\frac{2\omega Q(Mr-Q^2)}{M^2-Q^2}+\frac{3i M}{2Q},\quad
		h=\Rone+i\sig\lambda.
	\end{equation}
	
	The radial variables used below are
	\begin{equation}
		w^{(+)}=\Delta^{3/4}H_r^{-1}e^{i\xi_0r}z_2^{(+)},
		\,\,\,
		w^{(-)}=\Delta^{-3/4}H_r^{-1}e^{i\xi_0r}z_1^{(-)},
	\end{equation}
	where $z_{1,2}^{(\varsigma)}$ are the separated radial amplitudes defined in Ref. \cite{Hintz:2026jiw} and
	\begin{equation}
		H_r(r)=\exp\left[
		i\int_{2r_+}^{r}
		\frac{K(\tilde r)}{\Delta(\tilde r)}\,d\tilde r
		\right].
	\end{equation}
	For fixed $\varsigma$, we write $w=w^{(\varsigma)}$. The separated radial equations are then 
	\begin{align}\nonumber
		&\Biggl\{\left[\partial_r-\frac{(1+3\sig)\Delta'}{2\Delta}
		+\frac{2i\Kt}{\Delta}-\frac{(\Rone)'}{h}\right]
		\Delta\partial_r\\
		&\qquad-\bigl[(\Rone)^2+\lambda^2\bigr]\Biggr\}w=0,
		\label{eq:radial}
	\end{align}
	where primes in radial formulas denote $d/d r$.
	
	Multiplying by $h$ gives a numerically convenient form
	\begin{equation}
		A w''+B w'+Cw=0,
		\label{eq:cleared}
	\end{equation}
	with
	\begin{align}
		A&=\Delta h,\\
		B&=\frac{1-3\sig}{2}\Delta' h+2i\Kt h-\Delta(\Rone)',\\
		C&=-h\bigl[(\Rone)^2+\lambda^2\bigr].
		\label{eq:ABC}
	\end{align}
	This removes division by $h$ and thus produces an equivalent ordinary differential equation when $h\neq 0$.
	
	\noindent {\bf\emph{Angular regularity at the poles.}} 
	Let $z=\cos\theta$ and, for component $s=1,2$, define
	\begin{equation}
		\alpha_s=|\sig m+s|,\qquad \beta_s=|\sig m-s|.
	\end{equation}
	The regular component behavior used by the implementation is
	\begin{align}
		\chi_s(\theta\to0)&=\theta^{\alpha_s+1/2}
		\!\times\!\text{regular function of }\theta^2,
		\\
		\chi_s(\theta\to\pi)&=(\pi{-}\theta)^{\beta_s+1/2}
		\!\times\!\text{regular function of }(\pi{-}\theta)^2.
		\label{eq:poles}
	\end{align}
	The half-powers reflect the $\sqrt{\sin\theta}$ rescaling of the angular variables; they are not singular physical spin-weighted fields. Imposing both pole conditions selects discrete angular eigenvalues $\lambda=\lambda_{\ell m b}^{(\sig)}(\omega;a,Q,M)$, away from possible exceptional points. For more information, see Appendix \ref{app:angular-galerkin}.
	
	\noindent {\bf\emph{Ingoing behavior at the future event horizon.}} 
	For the physical perturbation, the horizon-ingoing phase is $e^{-i(\omega-m\Omega_H)r_*}$, with the appropriate field-dependent powers of $\Delta$, where
	\begin{equation}
		\Omega_H=\frac{a}{r_+^2+a^2},\qquad
		\frac{d r_*}{d r}=\frac{r^2+a^2}{\Delta}.
	\end{equation}
	The rescaling defining $w$ removes the ingoing phase and the associated horizon powers. The desired solution therefore has, generically, the regular expansion
	\begin{equation}
		w(r)=\sum_{j=0}^{\infty}b_j(r-r_+)^j.
		\label{eq:horizon}
	\end{equation}
	For $h(r_+)\neq0$, the substitution of $w\sim(r-r_+)^\alpha$ into the radial equation gives two Frobenius exponents: $\alpha=0$ and
	\begin{equation}
		\rho_H=\frac{1+3\varsigma}{2}
		-\frac{2iK(r_+)}{\Delta'(r_+)}.
	\end{equation}
	The transformation defining $w$ factors out the physical ingoing behavior at the event horizon. Consequently, the desired solution corresponds generically to the zero-exponent branch, represented by the regular expansion in Eq.~\eqref{eq:horizon}. The second local branch, proportional to $(r-r_+)^{\rho_H}$, remains a solution of the radial equation and is excluded by setting its coefficient to zero. Boundedness alone need not distinguish these branches for complex frequencies. At resonant frequencies, where the exponents differ by an integer, the ingoing solution requires a separate local analysis or identification by continuation.
	
	\noindent {\bf\emph{Outgoing behavior at infinity.}} 
	For $Q\omega\ne0$, the outgoing branch of the transformed radial equation has
	\begin{equation}
		w(r)\sim e^{\kappa r}r^p\sum_{j=0}^{\infty}a_jr^{-j},\quad
		\kappa=i(2\omega+\xi_0),\quad p=4i M\omega.
		\label{eq:infinity}
	\end{equation}
	The competing branch contains the exponential factor $e^{i\xi_0 r}$. The QNM boundary condition selects the outgoing solution by analytic continuation of its asymptotic behavior, rather than by requiring decay along the real radial axis. After extracting the outgoing prefactor, the competing branch contains $e^{-2i\omega r}$ multiplied by algebraic factors. For damped frequencies, $\operatorname{Im}\omega<0$, this contribution decays exponentially as $r\to+\infty$. Consequently, smoothness of the regularized radial function at compactified infinity does not, by itself, distinguish the outgoing solution from an admixture of the competing branch.
	
	After imposing boundary conditions on both the angular eigenvalue problem and the radial differential equation, we obtain a set of QNMs $\omega_n$, where $n$ is the overtone number. We have deliberately dropped the $\ell$ and $m$ indices for simplicity.
	
	\noindent {\bf\emph{Numerical method.}}
	The desired pair of equations is therefore described by the vanishing of some functional relations,
	$F_{\mathrm{ang}}(\omega,\lambda)=0,\,
	F_{\mathrm{rad}}(\omega,\lambda)=0$,
	where the first imposes angular regularity, while the second imposes radial QNM conditions. Separation eliminates the coupling between angular and radial functions, though it does not make their eigenvalues independent.
	
	The numerical implementation combines known spectral techniques, i.e., angular Galerkin projection and radial Chebyshev collocation. Spectral approaches have previously been used for angular eigenvalue problems in Kerr QNM calculations \cite{Cook:2014cta}, for radial QNM equations \cite{Jansen:2017oag}, and for the coupled gravitoelectric perturbation equations of KN BHs \cite{Dias:2015wqa,Dias:2022oqm,Pombo:2026xsj}. Here, we apply the combination of these techniques to the separated angular--radial system of Ref.~\cite{Hintz:2026jiw}, solving simultaneously for the complex frequency and separation constant.
	
	\noindent{\bf\emph{Convergence of the numerical method.}}
	The default numerical convergence test includes $(\omega,\lambda)$ calculations for various KN BHs with an increasing sequence of $(N,L)$, where $L$ is the angular basis cutoff and $N$ is the radial interpolation degree, corresponding to $N+1$ Chebyshev--Lobatto grid points, used to discretize the radial equation (see Appendix \ref{app:radial-angular discretization}). The sequence we use is $(N,L)=(32+16j,\max(\ell+4+2j,8+2j))$, $j=0,1,2$. Independent changes of $N$, $L$, quadrature order, and arithmetic precision have also been investigated. 
	
	All numerical tests that have been performed result in converging $\omega$ and $\lambda$ eigenvalues. The QNMs found have no discrepancies up to 9 decimal points for $b=+1$ and up to 11 decimal points for $b=-1$. Similar results apply for the $\lambda$ eigenvalues. For an illustration, see Table \ref{table:numerical convergence}, where we show QNMs and separation constants converging when the resolution $(N,L)$ increases.
	
	\begin{figure*}
		\centering
		\includegraphics[width=\textwidth]
		{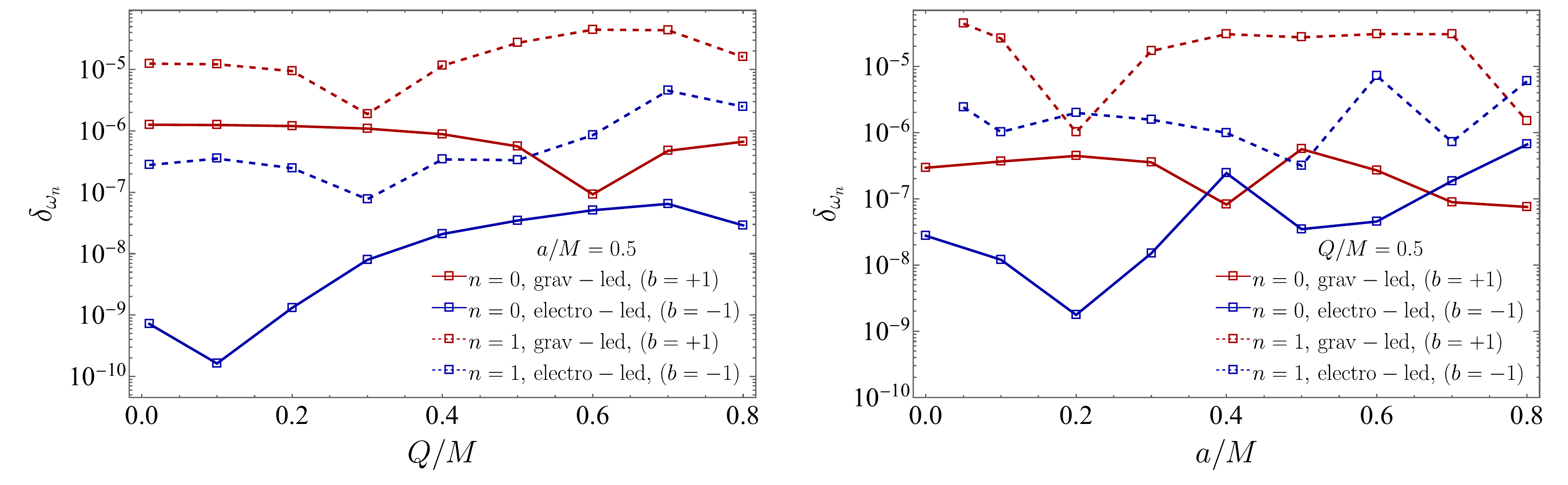}
		\caption{Absolute difference $\delta_{\omega_n}$ between the results from our separated numerical code and the data from a fully coupled solver, for the KN gravitational and electromagnetic-led QNMs with $\ell=m=2$, and spin label $\sig=1$.  On the left, we fix $a/M=0.5$ and vary $Q/M$, while on the right we fix $Q/M=0.5$ and vary the spin.}
		\label{fig:coupled_vs_decoupled}
	\end{figure*}
	
	\noindent {\bf\emph{Limiting cases.}}
	From now on, we employ the maximum of the above $(N,L)$ values. The checks in the previous section regard convergence of the numerical code. Convergence is a necessary diagnostic, but it does not provide a proof that all roots are physical QNMs. In this section we perform validity tests at limiting cases of the KN BH, such as the Schwarzschild, Reissner--Nordstr\"om (RN), and Kerr limits, as well as how well they match the known published literature, which uses a variety of numerical methods distinct from ours. We calculate fundamental QNMs for a wide variety of values in all limits. 
	
	In the Schwarzschild limit $a,Q\to 0$, the convergence analysis of Schwarzschild QNMs in the parameter space demonstrates that our numerical solver finds the fundamental QNMs, with respect to a matrix solver \cite{Lin:2016sch}, with at least 6 digits of precision, therefore establishing convergence between independent numerical methods. These results are shown in Table \ref{table:schwarzschild}, where we display arbitrary choices of QNMs in the parameter space for illustrative purposes. It becomes apparent that the branch number $b$ is fundamental in choosing the gravitational or electromagnetic sector of Schwarzschild QNMs. Specifically, for $b=+1$ the QNMs correspond to the gravitational sector with perturbation spin $s=\pm2$, while for $b=-1$ the QNMs correspond to the electromagnetic sector with perturbation spin $s=\pm1$.
	
	For RN QNMs, we compared our results (shown in Table~\ref{table:RN}) with those shown in Ref.~\cite{Kokkotas:1988fm}. We find agreement to all decimal points shown in that reference. The implication found for branch selection in Schwarzschild QNMs is demonstrated in the RN case also, where the $b=+1$ branch corresponds to gravitational-led QNMs, while the $b=-1$ branch corresponds to electromagnetic-led QNMs.
	
	For the Kerr limit, we take $Q\to 0$ and vary the spin. We compare our results with those shown in Refs.~\cite{Berti:2005ys,GRIT,CoG}. The agreement is around 6 decimal points, possibly due to the equations diverging at $Q=0$. Therefore, very small, but nonzero, BH charges have been used, with $10^{-4}\lesssim\frac{Q}{M}\lesssim 10^{-2}$.
	
	We explain the observations regarding different branches in Appendix \ref{app:Q-to-0} by demonstrating that a suitably rescaled version of the separated radial equations converges to the spin $2\varsigma=\pm 2$ (for $b=+1$) and $\varsigma=\pm 1$ (for $b=-1$) Teukolsky equations in the Kerr limit $Q\to 0$.
	
	\noindent{\bf\emph{QNMs: Separated equations vs.\ coupled system.}}
	We define the absolute difference 
	\begin{equation}
		\delta_{\omega_n}\equiv|M\omega_\textrm{sep}-M\omega_\textrm{coupled}|,
	\end{equation}
	where $\omega_\textrm{sep}$ are the QNMs calculated with the numerical method that uses the separated formulation \cite{Hintz:2026jiw}, while $\omega_\textrm{coupled}$ are the coupled gravitoelectric QNMs computed and tabulated in \cite{Pombo:2026xsj,pombo_2026_21359120}.
	
	Across the parameter ranges investigated, which we show in Fig. \ref{fig:coupled_vs_decoupled}, the absolute discrepancies range from approximately $10^{-10}$ to $10^{-6}$ for the fundamental modes shown, and reach a few times $10^{-5}$ for the first overtones. These differences reflect the combined numerical uncertainties of the two calculations.
	
	This agreement provides numerical validation of the separated formulation and of the identification of the computed roots as KN QNMs. The limitations of convergence do not seem to be an issue for near-extremal BHs. The agreement for overtones is less impressive; we cannot assess if this is a consequence of our numerical implementation or the results from the coupled system \cite{pombo_2026_21359120} that were used for the comparison.
	
	\noindent {\bf\emph{Spin-label isospectrality.}} 
	%
	\begin{figure*}
		\centering
		\includegraphics[width=\textwidth]
		{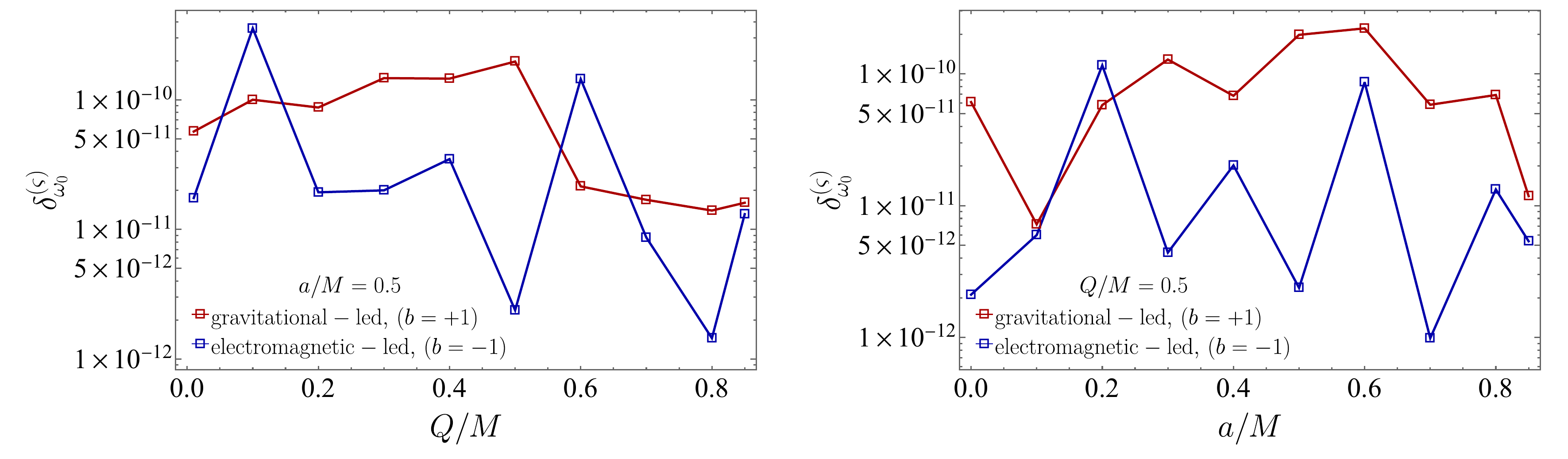}
		\caption{Absolute difference $\delta^{(\sig)}_{\omega_0}$ between different spin-label QNMs of fundamental KN gravitational- and electromagnetic-led perturbations with $\ell=m=2$. On the left, we fix $a/M=0.5$ and vary $Q/M$, and on the right we fix $Q/M=0.5$ and vary $a/M$.}
		\label{fig:decoupled_isospectrality}
	\end{figure*}
	KN \emph{parity}-isospectrality has, thus far, been supported only by perturbative and approximate numerical evidence. More specifically, Refs. \cite{Pani:2013ija,Pani:2013wsa} hint at parity-isospectrality in KN QNMs at the slow-rotation limit, to linear order in spin. Reference \cite{Blazquez-Salcedo:2022eik} discusses the slow-rotation limit and expands to second order in spin, and shows that isospectrality between axial- and polar-led perturbations for both electromagnetic-led and gravitational-led QNMs is approximately satisfied with good accuracy. A promising avenue towards an exact proof of parity-isospectrality of KN QNMs in the Dudley--Finley approximation has been presented in \cite{Silva:2025khf}. Finally, a general parity-isospectrality result has been reported in \cite{Weller:2026jqu}.
	
	We investigate here \emph{spin-label} isospectrality. This was studied in \cite{Mark:2014aja} in the small-charge limit. We numerically compare the fundamental frequencies obtained from the two spin-label descriptions, $\varsigma=\pm1$, for each angular branch. Their agreement provides a consistency check of the separated formulation and numerical implementation. We define
	\begin{equation}
		\delta^{(\sig)}_{\omega_0}\equiv |M\omega^{(+1)}-M\omega^{(-1)}| 
	\end{equation}
	for each perturbation branch. In Fig.~\ref{fig:decoupled_isospectrality} we show the agreement of the two spin-label descriptions ($\sig=\pm 1$) for a wide span of KN BH cases. Our fully numerical parameter-space sweep reveals a numerical isospectrality for spin label $\sig=+1$ and $\sig=-1$ KN fundamental QNMs, for each branch of our solutions. The absolute discrepancies range from roughly $10^{-12}$ to a few $10^{-10}$. These reflect the combined numerical uncertainties of the two calculations. The calculations also yield consistent spin-label results for the near-extremal example shown in Table \ref{table:isospectrality}.

	The KN spin-label isospectrality we describe here only gives numerical evidence, but does not constitute a general all-spin analytic proof. An analytic boundary-condition-preserving map would be needed for a general isospectrality statement. In the Kerr limit, the Teukolsky--Starobinsky identities \cite{Casals:2020fsb} imply the isospectrality of the spin $+2$ and spin $-2$ (and also of the spin $+1$ and spin $-1$) perturbation equations. Analogous analytic identities are, at present, not known for KN. We give a simple proof of spin-label isospectrality for the angular eigenvalue problem in Appendix~\ref{app:angular-basis}.
	
	\noindent{\bf\emph{Discussion and outlook.}}
	We have applied the separated KN perturbation equations to the computation of QNMs and their associated angular eigenvalues. Comparisons with independent calculations of the coupled gravitoelectric system show agreement for the fundamental modes and first overtones investigated. This is demonstrated in Fig. \ref{fig:coupled_vs_decoupled}. Together with the limiting-case checks and the vanishing-charge reduction to the Teukolsky equations, these results support the separated formulation as a practical approach to KN spectroscopy. We also find numerical evidence consistent with spin-label isospectrality: for each of the gravitational-led and electromagnetic-led branches, the fundamental frequencies obtained with $\varsigma=\pm1$ agree to within approximately $10^{-12}$--$10^{-10}$ in $M\omega$ across the configurations shown in Fig. \ref{fig:decoupled_isospectrality}. This agreement motivates an analytical relation between the two spin descriptions, while its connection to parity isospectrality remains to be established.
	
	The separated equations open several directions for further study, including higher overtones, near-horizon mode families, and the approach to extremality. The formulation at vanishing charge developed here also offers a route to calculations spanning the Kerr limit without singular coefficients. Independent implementations of the outgoing boundary condition would strengthen numerical validation, particularly for strongly damped modes. Furthermore, the zero-frequency separated equations provide a starting point for determining horizon-regular solutions in the presence of external gravitational and electromagnetic tidal fields. Matching their asymptotic behavior to the applied fields and induced multipole moments could give access to both the individual and mixed tidal responses. This requires relating the separated variables to physical perturbations and replacing the outgoing condition at infinity by the appropriate tidal asymptotics. Stationary, axisymmetric perturbations offer a natural first setting for this investigation. An explicit relation between the two spin descriptions, preserving the QNM boundary conditions, would further clarify the observed spectral agreement and its connection to physical parity sectors.
	
	\begin{acknowledgements}
		%
		KD acknowledges financial support provided by FCT–Fundação para a Ciência e a Tecnologia, I.P., under the Scientific Employment Stimulus – Individual Call – Grant No. \href{https://sciproj.ptcris.pt/12680EEC}{2023.07417.CEECIND/CP2830/CT0008}. KD also acknowledges funding from the European Union's Horizon Europe research and innovation programme under the Marie Skłodowska-Curie Actions project ``EinsteinWaves,'' Grant Agreement No.~101131233.
		The Center of Gravity is a Center of Excellence funded by the Danish National Research Foundation under grant No. DNRF184. We acknowledge support by VILLUM Foundation (grant no. VIL37766). V.C. is a Villum Investigator. V.C. acknowledges financial support provided under the European Union’s H2020 ERC Advanced Grant “Black holes: gravitational engines of discovery” grant agreement no. Gravitas–101052587. Views and opinions expressed are however those of the author only and do not necessarily reflect those of the European Union or the European Research Council. Neither the European Union nor the granting authority can be held responsible for them. This project has received funding from the European Union’s Horizon 2020 research and innovation programme under the Marie Sklodowska-Curie grant agreement No 101007855 and No 101131233. 
		This work is supported by Simons Foundation International \cite{sfi} and the Simons Foundation \cite{sf} through Simons Foundation grant SFI-MPS-BH-00012593-11.
		PH acknowledges support from the U.S.\ National Science Foundation under grant DMS-2554160.
		We acknowledge financial support from FCT for the Center for Astrophysics and Gravitation (CENTRA/IST/ULisboa) through Grants No.~\href{https://doi.org/10.54499/UID/PRR/00099/2025}{UID/PRR/00099/2025} and No.~\href{https://doi.org/10.54499/UID/00099/2025}{UID/00099/2025}.
	\end{acknowledgements}
	
	\clearpage
	\begin{center}
		{\bf\large Supplemental material}
	\end{center}
	\appendix
	
	\section{Angular operator and complex frequencies}\label{app:angular-basis}
	
	Here, we relate the two-component angular equation used in our numerical calculation to the angular operator of Ref.~\cite{Hintz:2026jiw}. We introduce the notation
	\begin{align}
		\begin{split}
			W(\theta)
			&:= \Theta_0^{(\varsigma)}(\theta) \\
			&= \frac{3}{2}\cot\theta
			+ \varsigma\left[
			\frac{m}{\sin\theta}
			- \frac{a\omega(M^2+Q^2)}{M^2-Q^2}\sin\theta
			\right],
		\end{split}
		\label{eq:angular-W}
		\\
		T(\theta)
		&:= \Theta_1^{(\varsigma)}(\theta)
		= -\frac{3M}{2Q}
		- \varsigma\frac{2a\omega MQ}{M^2-Q^2}\cos\theta,
		\label{eq:angular-T}
	\end{align}
	where $\varsigma=\pm1$ and $Q\neq0$.
	
	As in Ref.~\cite{Hintz:2026jiw}, we restrict the angular operator denoted $\mathcal{A}^{(\varsigma)}$ there to $\ker(\mathcal{S}-i)$, where $\mathcal{S}$ denotes the symmetry operator defined in \cite{Hintz:2026jiw}, and $\ker(\mathcal{S}-i)=\{x\in\mathbb{C}^4\colon\mathcal{S}x-i x=0\}$. In the basis
	\begin{equation}
		e_1=(i,1,0,0)^T,
		\qquad
		e_2=(0,0,-i,1)^T,
	\end{equation}
	the restricted eigenvalue problem takes the form
	\begin{equation}
		\mathcal{A}u=\lambda u,
		\qquad
		\mathcal{A}=
		\begin{pmatrix}
			i\partial_\theta & T-iW\\
			T+iW & -i\partial_\theta
		\end{pmatrix},
		\label{eq:angular-original-basis}
	\end{equation}
	where $u=(u_1,u_2)^T$ contains the coefficients in this basis.
	
	The explicit calculation in Lemma~2 of Ref.~\cite{Hintz:2026jiw} concerns the specialization $\omega=0$, for which
	\begin{equation}
		W(\theta)\big|_{\omega=0}
		= \frac{3}{2}\cot\theta+\frac{\varsigma m}{\sin\theta},
		\quad
		T(\theta)\big|_{\omega=0}
		= -\frac{3M}{2Q}.
	\end{equation}
	
	For the QNM calculation, we retain the full frequency dependence in Eqs.~\eqref{eq:angular-W} and \eqref{eq:angular-T} and consider complex $\omega$. To obtain the representation used numerically, we introduce the constant change of basis
	\begin{equation}
		u=H\chi,
		\quad
		H=\frac{1}{\sqrt{2}}
		\begin{pmatrix}
			1&1\\
			1&-1
		\end{pmatrix},
		\quad
		H^{-1}=H^\dagger=H.
		\label{eq:angular-basis-change}
	\end{equation}
	Equivalently,
	\begin{equation}
		u_1=\frac{\chi_1+\chi_2}{\sqrt{2}},
		\qquad
		u_2=\frac{\chi_1-\chi_2}{\sqrt{2}}.
	\end{equation}
	Multiplying Eq.~\eqref{eq:angular-original-basis} by $H$ gives
	\begin{equation}
		\mathcal{B}\begin{pmatrix}
			\chi_1\\
			\chi_2
		\end{pmatrix}
		=
		\lambda
		\begin{pmatrix}
			\chi_1\\
			\chi_2
		\end{pmatrix},
		\label{eq:angular-numerical-basis}
	\end{equation}
	where
	\begin{equation}
		\mathcal{B}\equiv\begin{pmatrix}
			T & i(\partial_\theta+W)\\
			i(\partial_\theta-W) & -T
		\end{pmatrix}.
		\label{eq:angular-B}
	\end{equation}
	Since $H$ is independent of $\theta$, this transformation introduces no additional derivative terms. It preserves the eigenvalue $\lambda$ and maps the angular regularity conditions to the corresponding conditions on $\chi$. Thus, Eq.~\eqref{eq:angular-numerical-basis} is an exact change of representation of the restricted angular operator, rather than an approximation. This algebraic equivalence holds for complex $\omega$, although self-adjointness of the real-frequency angular problem does not generally extend to complex frequencies.
	
	Defining for $\chi=(\chi_1,\chi_2)^T$
	\begin{equation}
		(U\chi)(\theta) = \begin{pmatrix} \chi_1(\pi-\theta) \\ -\chi_2(\pi-\theta) \end{pmatrix},
	\end{equation}
	and restoring the spin label and angular momentum label for the map $\mathcal{B}^{(\varsigma)}(\omega)$ from Eq.~\eqref{eq:angular-B}, we have
	\begin{equation}
		\mathcal{B}^{(-)}(\omega) U = U\mathcal{B}^{(+)}(\omega).
	\end{equation}
	Since $U$ preserves the regularity conditions at the poles, this proves spin-label isospectrality for the angular eigenvalue problem.
	
	\section{The vanishing charge limit}
	\label{app:Q-to-0}
	
	Motivated by \eqref{eq:spherical-lambda}, we seek the angular eigenvalues in the form
	\begin{equation}
		\lambda = b\Bigl(\frac{3 M}{2 Q}+\frac{Q}{M}\tilde\lambda\Bigr).
	\end{equation}
	The limiting reduction follows branches for which $\tilde{\lambda}$ remains finite as $Q\to0$. For $b=+1$, resp.\ $b=-1$, we seek the eigenfunctions of $\mathcal{B}$ in Eq.~\eqref{eq:angular-B} in the form $(\frac{Q}{M}\tilde\chi_1(\theta),\chi_2(\theta))$, resp.\ $(\chi_1(\theta),\frac{Q}{M}\tilde\chi_2(\theta))$, which leads to the generalized eigenvalue problems
	\begin{equation}
		\label{eq:ang+1}
		\begin{pmatrix}
			-3+\frac{Q^2}{M^2}\tilde T & i(\partial_\theta+W) \\
			i(\partial_\theta-W) & -\tilde T
		\end{pmatrix}
		\begin{pmatrix} \tilde\chi_1 \\ \chi_2 \end{pmatrix}
		=
		\tilde\lambda \begin{pmatrix} \frac{Q^2}{M^2} & 0 \\ 0 & 1 \end{pmatrix} \begin{pmatrix} \tilde\chi_1 \\ \chi_2 \end{pmatrix},
	\end{equation}
	and
	\begin{equation}
		\label{eq:ang-1}
		\begin{pmatrix}
			\tilde T & i(\partial_\theta+W) \\
			i(\partial_\theta-W) & 3-\frac{Q^2}{M^2}\tilde T
		\end{pmatrix}
		\begin{pmatrix} \chi_1 \\ \tilde\chi_2 \end{pmatrix}
		= -\tilde\lambda\begin{pmatrix} 1 & 0 \\ 0 & \frac{Q^2}{M^2} \end{pmatrix}
		\begin{pmatrix} \chi_1 \\ \tilde\chi_2 \end{pmatrix},
	\end{equation}
	where we set
	\begin{equation}
		\tilde T(\theta) := -\varsigma\frac{2 a\omega M^2\cos\theta}{M^2-Q^2}.
	\end{equation}
	Note that for fixed $Q\neq 0$, both of these generalized eigenvalue problems encode the \emph{full} angular spectrum, merely with different parameterizations of the eigenvalues. In the limit $Q\to 0$, the two systems can no longer be related by a finite shift of $\tilde\lambda$. Importantly, they remain regular in this limit.
	
	After angular separation, consider the radial ordinary differential equation system, Eq.~(11) in \cite{Hintz:2026jiw}, which involves the function $R_1^{(\varsigma)}=i\frac{3 M}{2 Q}+\frac{Q}{M}\tilde R_1^{(\varsigma)}$ where
	\begin{equation}
		\tilde R_1^{(\varsigma)} = -\varsigma\frac{2\omega M(M r-Q^2)}{M^2-Q^2}.
	\end{equation}
	Since $R_1^{(\varsigma)}-i\lambda=\mathcal{O}(\frac{Q}{M})$ for $b=+1$, while $R_1^{(\varsigma)}+i\lambda=\mathcal{O}(\frac{Q}{M})$ for $b=-1$, we must weight $z_{1/2}^{(\pm)}$ by different powers of $\frac{Q}{M}$ depending on the choice of $b$. Concretely, for $b=+1$ we write $z_1^{(\varsigma)}:=\frac{Q}{M}\tilde z_1^{(\varsigma)}$, $z_2^{(\varsigma)}=\tilde z_2^{(\varsigma)}$, while for $b=-1$ we write $z_1^{(\varsigma)}=\tilde z_1^{(\varsigma)}$, $z_2^{(\varsigma)}=\frac{Q}{M}\tilde z_2^{(\varsigma)}$. The resulting systems are
	\begin{align}
		(\partial_r-\Delta^{-\frac12}R_0^{(\varsigma)})\tilde z_1^{(\varsigma)} &= \Delta^{-\frac12}(\tilde R_1^{(\varsigma)}-i\tilde\lambda)\tilde z_2^{(\varsigma)}, \\
		(\partial_r+\Delta^{-\frac12}R_0^{(\varsigma)})\tilde z_2^{(\varsigma)} &= \Bigl(\frac{3 i}{\Delta^{\frac12}} + \frac{Q^2}{M^2\Delta^{\frac12}}(\tilde R_1^{(\varsigma)}+i\tilde\lambda)\Bigr)\tilde z_1^{(\varsigma)}
	\end{align}
	for $b=+1$ and
	\begin{align}
		(\partial_r-\Delta^{-\frac12}R_0^{(\varsigma)})\tilde z_1^{(\varsigma)} &= \Bigl(\frac{3 i}{\Delta^{\frac12}} + \frac{Q^2}{M^2\Delta^{\frac12}}(\tilde R_1^{(\varsigma)}+i\tilde\lambda)\Bigr)\tilde z_2^{(\varsigma)}, \\
		(\partial_r+\Delta^{-\frac12}R_0^{(\varsigma)})\tilde z_2^{(\varsigma)} &= \Delta^{-\frac12}(\tilde R_1^{(\varsigma)}-i\tilde\lambda)\tilde z_1^{(\varsigma)}
	\end{align}
	for $b=-1$.
	
	We next pass to horizon-regular quantities: for $b=+1$ we put
	\begin{equation}
		\tilde w^{(\varsigma)} = \Delta^{\varsigma-\frac14}H_r^{-1}e^{i\xi_0 r}\tilde z_2^{(\varsigma)}, \ 
		\tilde w_\sharp^{(\varsigma)} = \Delta^{\frac{\varsigma}{2}-\frac14}H_r^{-1}e^{i\xi_0 r}\tilde z_1^{(\varsigma)},
	\end{equation}
	which, after undoing the previous rescalings by $\frac{Q}{M}$, matches Eq.~(14) in \cite{Hintz:2026jiw} when $\varsigma=+1$, and switches $\tilde w$ and $\tilde w_\sharp$ when $\varsigma=-1$; for $b=-1$ on the other hand we put
	\begin{equation}
		\tilde w^{(\varsigma)}=\Delta^{\frac{\varsigma}{2}-\frac14}H_r^{-1}e^{i\xi_0 r}\tilde z_1^{(\varsigma)},\ 
		\tilde w_\sharp^{(\varsigma)} = \Delta^{\varsigma-\frac14}H_r^{-1}e^{i\xi_0 r}\tilde z_2^{(\varsigma)},
	\end{equation}
	which matches Eq.~(14) in \cite{Hintz:2026jiw} when $\varsigma=-1$, and switches $\tilde w$ and $\tilde w_\sharp$ when $\varsigma=+1$. For all choices of $b=\pm 1$ and for $\varsigma=\pm 1$, the resulting first-order system can be written in the form
	\begin{align}
		\begin{split}
			&\Bigl[\Delta^{\frac{1-\varsigma b}{2}}\partial_r+\frac{1-\varsigma b}{2}\Bigl(\frac{1-3\varsigma}{2}\Delta'+2 i\Kt\Bigr) \Bigr]\tilde w^{(\varsigma)} \\
			&\quad = \Bigl(3 i+\frac{Q^2}{M^2}(\tilde R_1^{(\varsigma)}+i\tilde\lambda)\Bigr)\tilde w_\sharp^{(\varsigma)},
		\end{split} \\
		\begin{split}
			&\Bigl[\Delta^{\frac{1+\varsigma b}{2}}\partial_r+\frac{1+\varsigma b}{2}\Bigl(\frac{1-3\varsigma}{2}\Delta'+2 i\Kt\Bigr)\Bigr]\tilde w_\sharp^{(\varsigma)} \\
			&\quad = (\tilde R_1^{(\varsigma)}-i\tilde\lambda)\tilde w^{(\varsigma)}.
		\end{split}
	\end{align}
	
	Solving the first equation for $\tilde w_\sharp^{(\varsigma)}$ (assuming there is no division by $0$) and substituting into the second equation, one obtains the radial master equation
	\begin{equation}
		\mathcal{L}_b^{(\varsigma)}\tilde w^{(\varsigma)}=0,
	\end{equation}
	where, upon setting $c_b:=\frac{3+b}{4}$,
	\begin{equation}
		\begin{split}
			\mathcal{L}_b^{(\varsigma)} &= \Bigl(\partial_r - b\tfrac{(1+\varsigma b)(c_b\Delta'-i b\Kt)}{\Delta} - \tfrac{\frac{Q^2}{M^2}\partial_r\tilde R_1^{(\varsigma)}}{3 i+\frac{Q^2}{M^2}(\tilde R_1^{(\varsigma)}+i\tilde\lambda)}\Bigr) \\
			&\quad \circ \bigl(\Delta\partial_r + (1-\varsigma b)(b c_b\Delta'+i\Kt)\bigr) - 3 i(\tilde R_1^{(\varsigma)}-i\tilde\lambda) \\
			&\quad - \frac{Q^2}{M^2}\bigl((\tilde R_1^{(\varsigma)})^2+\tilde\lambda^2\bigr).
		\end{split}
	\end{equation}
	Unlike the radial master equation in \cite{Hintz:2026jiw}, these four separated equations (one for each pair of $(b,\varsigma)$) remain regular as $Q\to 0$. Concretely, at $Q=0$ we have $\tilde R_1^{(\varsigma)}=-2\varsigma\omega r$ and $\Kt=K$, and the radial master equation for $b=+1$ becomes
	\begin{equation}
		\begin{split}
			&\Bigl(\Delta\partial_r^2+\bigl((1-2\varsigma)\Delta'+2 i K\bigr)\partial_r \\
			&\quad +2(1-\varsigma)-3\tilde\lambda+2 i(4\varsigma-1)\omega r\Bigr)\tilde w^{(\varsigma)} = 0.
		\end{split}
	\end{equation}
	These are the (horizon-normalized) radial Teukolsky equations on Kerr for spin $2\varsigma=\pm 2$. For $b=-1$ on the other hand, the $Q\to 0$ limit of the radial master equation is
	\begin{equation}
		\begin{split}
			&\Bigl(\Delta\partial_r^2 + \bigl((1-\varsigma)\Delta'+2 i K\bigr)\partial_r \\
			&\quad - (1+\varsigma)-3\tilde\lambda+2 i(2\varsigma-1)\omega r\Bigr)\tilde w^{(\varsigma)} = 0.
		\end{split}
	\end{equation}
	
	\emph{These are the (horizon-normalized) radial Teukolsky equations for spin $\varsigma=\pm 1$.} For $s=2\varsigma$ (when $b=+1$) and $s=\varsigma$ (when $b=-1$), the substitution $R_s=H_r\Delta^{-s}\tilde w$ gives the usual form of the Teukolsky equations. The usual angular separation constant $\Lambda_s={}_s A_{l m}+a^2\omega^2-2 a m\omega$ is $3\tilde\lambda-2-2\varsigma$ for $b=+1$ and $3\tilde\lambda+1-\varsigma$ for $b=-1$. This explains why the QNMs we find numerically for $b=+1$ are the gravitational-led modes, and those for $b=-1$ the electromagnetic-led modes.
	
	\section{Numerical discretization}\label{app:radial-angular discretization}
	\subsection{Angular Galerkin discretization}
	\label{app:angular-galerkin}
	
	We discretize the two-component angular eigenvalue problem in Eq. \eqref{eq:angular-numerical-basis} using a Galerkin expansion whose basis functions incorporate regularity at both angular poles.
	
	For each component $s=1,2$, we introduce
	\begin{align*}
		z&=\cos\theta,
		\qquad\,\,\,\,\,\,\,\,\,\,
		\alpha_s=|\varsigma m+s|,\\
		\beta_s&=|\varsigma m-s|,
		\quad
		j_{\min,s}=\max(|m|,s).
	\end{align*}
	The truncated expansion of angular eigenstates has the form
	\begin{equation}
		\chi_s(\theta)\approx
		\sum_{j=j_{\min,s}}^{L}
		c_{js}\,\Phi_{js}(\cos\theta),
		\label{eq:angular-expansion}
	\end{equation}
	where $c_{js}$ are complex coefficients and $L$ is the angular cutoff. Increasing $L$ tests whether the angular truncation is sufficiently accurate. Increasing $L$, therefore, improves the representation of a chosen mode. The basis functions are
	\begin{equation}
		\Phi_{js}(z)=
		\mathcal{N}_{js}
		(1-z)^{\alpha_s/2+1/4}
		(1+z)^{\beta_s/2+1/4}
		P_{j-j_{\min,s}}^{(\alpha_s,\beta_s)}(z),
		\label{eq:angular-jacobi-basis}
	\end{equation}
	with $P_n^{(\alpha,\beta)}$ denoting a Jacobi polynomial. 
	
	The positive normalization constant $\mathcal{N}_{js}$ is chosen so that
	\begin{equation}
		\int_0^\pi
		|\Phi_{js}(\cos\theta)|^2\,d\theta=1.
	\end{equation}
	Explicitly, with $n=j-j_{\min,s}$, we have
	\begin{equation}
		\mathcal{N}_{js}
		=
		\left[
		\frac{
			(2n+\alpha_s+\beta_s+1)\,n!\,
			\Gamma(n+\alpha_s+\beta_s+1)
		}{
			2^{\alpha_s+\beta_s+1}
			\Gamma(n+\alpha_s+1)\,
			\Gamma(n+\beta_s+1)
		}
		\right]^{1/2}.
	\end{equation}
	The summation index $j$ labels basis functions, which are different from the mode label $\ell$. Likewise, the component index $s$ is distinct from $\varsigma$.
	
	The endpoint factors in Eq.~\eqref{eq:angular-jacobi-basis} impose the required pole behavior. Since
	\begin{equation}
		1-\cos\theta\sim\frac{\theta^2}{2},
		\qquad
		1+\cos\theta\sim\frac{(\pi-\theta)^2}{2}
	\end{equation}
	near the north and south poles, respectively, the expansion has the form
	\begin{align}
		\chi_s(\theta\to0)
		&=\theta^{\alpha_s+1/2}
		\!\times\!\text{regular function of }\theta^2,
		\\
		\chi_s(\theta\to\pi)
		&=(\pi{-}\theta)^{\beta_s+1/2}
		\!\times\!\text{regular function of }(\pi{-}\theta)^2.
	\end{align}
	Thus, angular regularity is built into the approximation space. The half-integer powers arise from the rescaling used to define the angular variables and do not imply singular physical perturbations.
	
	The scalar inner product is
	\begin{equation}
		\langle f,g\rangle
		=\int_0^\pi f^*(\theta)g(\theta)\,d\theta
		=\int_{-1}^{1}
		\frac{f^*(z)g(z)}{\sqrt{1-z^2}}\,dz.
		\label{eq:angular-inner-product}
	\end{equation}
	Combining this measure with the endpoint factors yields the Jacobi weight:
	\begin{equation}
		\frac{
			(1-z)^{\alpha_s+1/2}(1+z)^{\beta_s+1/2}
		}{\sqrt{1-z^2}}
		=(1-z)^{\alpha_s}(1+z)^{\beta_s}.
	\end{equation}
	The normalization constants $\mathcal{N}_{js}$ are therefore chosen so that
	\begin{equation}
		\langle\Phi_{js},\Phi_{ks}\rangle=\delta_{jk}.
	\end{equation}
	
	To formulate the projection, we define the vector basis
	\begin{equation}
		\Psi_{j1}=
		\begin{pmatrix}\Phi_{j1}\\0\end{pmatrix},
		\qquad
		\Psi_{j2}=
		\begin{pmatrix}0\\\Phi_{j2}\end{pmatrix}.
	\end{equation}
	The vector inner product is the sum of the scalar inner products of the two components. Substituting the truncated expansion into the differential equation defines the residual
	\begin{equation}
		\mathcal{E}=\mathcal{B}(\omega)\chi-\lambda\chi.
	\end{equation}
	The Galerkin prescription requires
	\begin{equation}
		\langle\Psi_{js},\mathcal{E}\rangle=0,
	\end{equation}
	for every retained basis vector. Thus, the residual is orthogonal to the finite-dimensional approximation space, rather than being required to vanish pointwise.
	
	Because $T$ and $W$ depend linearly on $\omega$, we can write
	\begin{equation}
		\mathcal{B}(\omega)=\mathcal{B}_0+\omega\mathcal{B}_1.
	\end{equation}
	For a fixed trial frequency $\omega$, the projected angular problem is a matrix eigenvalue problem for $\lambda$:
	\begin{equation}
		\bigl(G_0+\omega G_1-\lambda I\bigr)c=0,
		\label{eq:angular-galerkin-matrix}
	\end{equation}
	where $c$ collects the coefficients of both components, and
	\begin{equation}
		(G_p)_{js,kt}
		=\langle\Psi_{js},\mathcal{B}_p\Psi_{kt}\rangle,
		\qquad p=0,1.
	\end{equation}
	
	For example, the upper-right block of the projected angular operator contains the matrix elements
	\begin{equation}
		\bigl[G_0+\omega G_1\bigr]_{j1,k2}
		=
		\int_0^\pi
		\Phi_{j1}^{*}\,
		i(\partial_\theta+W)\Phi_{k2}\,d\theta.
	\end{equation}
	Although the matrix depends linearly on $\omega$, its eigenvalues $\lambda(\omega)$ generally do not.
	
	In the numerical implementation, the matrix elements are evaluated using Gauss--Legendre quadrature with $4L+16$ nodes in $z$, including the factor $(1-z^2)^{-1/2}$ from Eq.~\eqref{eq:angular-inner-product}. Derivatives of the basis functions are evaluated analytically. The quadrature points serve to approximate the projection integrals as integration points, i.e., they are not collocation points at which the angular differential equation is imposed.
	
	The total matrix dimension is
	\begin{equation}
		N_{\mathrm{ang}}
		=\sum_{s=1}^{2}\bigl(L-j_{\min,s}+1\bigr).
	\end{equation}
	For example, $m=2$ and $L=8$ give seven basis functions per component and a $14\times14$ matrix. When $|m|<2$, the lower basis functions allowed by $j_{\min,s}$ must also be retained, even when the target mode has $\ell\ge2$.
	
	At each trial frequency, the desired angular eigenvalue branch is identified by continuation from its spherical-limit label $\ell$ and eigenvalue sign. The angular and radial problems are then solved for a common pair $(\omega,\lambda)$. Increasing $L$ tests angular convergence while keeping the target branch fixed. 
	
	\subsection{Radial compactification and prefactor}
	For the radial compactification we set
	\begin{equation}
		x=1-\frac{r_+}{r},\quad
		w=F(r)f(x),\quad
		F(r)=e^{\kappa(r-r_+)}\left(\frac{r}{r_+}\right)^p.
	\end{equation}
	Then $x=0$ is the horizon and $x=1$ represents infinity. We then define
	\begin{equation}
		t=\frac{d x}{d r}=\frac{r_+}{r^2},\qquad
		g=\frac{F'}{F}=\kappa+\frac{p}{r}.
	\end{equation}
	The full chain rule is
	\begin{equation}
		\frac{w''}{F}=t^2f_{xx}+(t'+2gt)f_x+(g'+g^2)f.
		\label{eq:chain}
	\end{equation}
	In particular, the coefficient of $f_x$ contains $t'$, not $tt'$. After multiplying the transformed equation by $1-x$, we can write
	\begin{equation}
		c_2 f_{xx}+c_1f_x+c_0f=0,
		\label{eq:compact}
	\end{equation}
	where
	\begin{align}
		c_2&=(1-x)At^2,\\
		c_1&=(1-x)\left[A(t'+2gt)+Bt\right],\\
		c_0&=(1-x)\left[A(g'+g^2)+Bg+C\right].
		\label{eq:compact-coefficients}
	\end{align}
	Here $A(r)$, $B(r)$, and $C(r)$ are the coefficient functions of the radial equation \eqref{eq:cleared}. In the compactified equation, all radial functions are evaluated at $r=r_+/(1-x)$. Leading large-$r$ terms cancel for the choices of $\kappa$ and $p$ in Eq.~\eqref{eq:infinity}. The numerical code assembles the resulting finite polynomials before evaluating endpoints, avoiding subtraction of divergent quantities.
	
	\subsection{Chebyshev collocation and endpoint rows}
	At $N+1$ Chebyshev--Lobatto grid points
	\begin{equation}
		x_j=\frac{1-\cos(j\pi/N)}2,\qquad j=0,\ldots,N,
	\end{equation}
	let $\bm D$ be the differentiation matrix with respect to $x$. The radial matrix is
	\begin{equation}
		\bm R(\omega,\lambda)
		=\operatorname{diag}(c_2)\bm D^2
		+\operatorname{diag}(c_1)\bm D
		+\operatorname{diag}(c_0).
	\end{equation}
	The regularized differential equation is imposed at all collocation nodes, including the endpoints through the finite limits of its coefficients. Generically, $c_2(0)=c_2(1)=0$, so the endpoint rows reduce to limiting first-order relations. These rows enforce compatibility with the differential equation at the boundaries but they do not independently establish the selection of the physical QNM branches. In particular, at infinity the competing solution can be smooth and exponentially small after removal of the outgoing prefactor. The outgoing interpretation must, therefore, be assessed through convergence studies and independent validation of the resulting frequencies.
	
	\section{Joint eigenvalue determination}
	\label{app:joint-eigenvalues}
	
	The angular and radial discretizations yield
	\begin{equation}
		G(\omega,\lambda)c=0,
		\qquad
		R(\omega,\lambda)f=0,
	\end{equation}
	where $c$ contains the angular expansion coefficients and $f$ contains the radial values at the collocation points. The desired pair $(\omega,\lambda)$ must allow nontrivial solutions of both systems.
	
	We determine the common eigenvalues using normalized scalar residuals, following the general approach of bordered formulations and implicit-determinant methods for nonlinear eigenvalue problems \cite{Spence:2005,Guttel_Tisseur_2017}. The construction below applies this approach separately to the angular and radial matrices and solves the two resulting conditions simultaneously.
	
	To express this requirement as two scalar conditions, consider either matrix $\mathcal{M}=G$ or $R$. The homogeneous equations determine each eigenvector only up to an overall complex amplitude. We fix this freedom by setting one non-zero component to unity, i.e., $v_k=1$. This normalization excludes the trivial solution without changing the eigenvalues or imposing an additional physical boundary condition. After fixed row scaling, one row is replaced by a normalization condition $v_k=1$. Solving the modified linear system determines a candidate vector $v$. The residual of the omitted row defines a complex function $F_{\mathcal{M}}(\omega,\lambda)$. Whenever the modified matrix is invertible, a zero of this residual restores the omitted equation and gives a non-trivial null vector of $\mathcal{M}$. The replacement row and normalization component are selected using the singular vectors associated with the smallest singular value at the initial estimate and are held fixed during each local search.
	
	The joint spectral conditions are, thus,
	\begin{equation}
		F_G(\omega,\lambda)=0,
		\qquad
		F_R(\omega,\lambda)=0.
	\end{equation}
	Their real and imaginary parts give four real equations for $\operatorname{Re}\omega$, $\operatorname{Im}\omega$, $\operatorname{Re}\lambda$, and $\operatorname{Im}\lambda$. These equations are solved using bounded nonlinear least squares followed by a local root refinement, with residual scaling based on their sensitivities at the initial estimate.
	
	The angular branch is initialized at $\omega=0$ using Eq. \eqref{eq:spherical-lambda}. It is followed to the trial frequency along $\omega\mapsto\tau\omega$, $0\le\tau\le1$, selecting successive eigenvectors by their largest normalized overlap. The resulting eigenvalue supplies the angular initial estimate, and the branch assignment is checked again at the converged frequency.
	
	When varying the BH parameters, previously computed frequencies provide initial estimates at neighboring points, while the angular branch is tracked at each new background. This continuation helps track the same mode as the BH parameters vary, reducing the risk of converging to a different branch or overtone. The radial overtone index requires identification from a known mode or an independent reference; it is not fixed by the angular labels alone. Candidate solutions are subsequently subjected to the convergence and validation checks described in the main text.
	
	\section{Tabulated QNMs}
	
	\begin{table}[ht]
		\centering
		\caption{Illustrative numerical convergence of our QNM implementation that uses the decoupled equations \eqref{eq:angular-two}, \eqref{eq:radial} for a KN BH with $M=1$, $a=0.2$, $Q=0.4$, $m=\ell=2$, $\sig=+1$, and $b=\pm1$.}
		\label{table:numerical convergence}
		\resizebox{\columnwidth}{!}{%
			\begin{tabular}{|c|c|c|}
				\toprule
				$b$ & $M\omega_0$ & $\lambda$\\\hline
				\multicolumn{3}{|c|}{ $(N,L)=(32,8)$ } \\\hline
				$+1$ & $0.4099182-0.088613i$ & $4.1671483+0.017881i$\\
				$-1$ & $0.51504862-0.0955714i$ & $-4.1796762-0.01280045i$\\\hline\hline
				
				\multicolumn{3}{|c|}{ $(N,L)=(48,10)$ } \\\hline
				$+1$ & $0.409918268-0.088613263i$ & $4.1671482212+0.017880575i$\\
				$-1$ & $0.51504859138-0.09557140698i$ & $-4.17967620174-0.01280045060i$\\\hline\hline
				
				\multicolumn{3}{|c|}{$(N,L)=(64,12)$ } \\\hline
				$+1$ & $0.409918268-0.088613263i$ & $4.1671482212+0.017880575i$\\
				$-1$ & $0.51504859138-0.09557140698i$ & $-4.17967620174-0.01280045060i$\\\hline\hline
		\end{tabular}}
	\end{table}
	
	\begin{table}[H]
		\centering
		\caption{QNMs at the Schwarzschild limit, where $a=0$ and $Q\to0$ of our separated numerical code with $m=1$, for $\ell=2$ and $\ell=3$. We compare our QNMs with a code developed based on a matrix method in Ref. \cite{Lin:2016sch}.}
		\label{table:schwarzschild}
		\resizebox{\columnwidth}{!}{%
			\begin{tabular}{|c|c||c|c|}
				\hline
				\multicolumn{2}{|c||}{$\ell=2$} &
				\multicolumn{2}{c|}{$\ell=3$} \\\hline
				$b$ & $M\omega$ & $b$ & $M\omega$ \\\hline
				
				\multicolumn{2}{|c||}{$\mathrm{Separate\ code},\,\sig=\pm1$} &
				\multicolumn{2}{c|}{$\mathrm{Separate\ code},\,\sig=\pm1$} \\\hline
				$-1$ & $0.45759552 - 0.09500443 i$ &
				$-1$ & $0.656899368 - 0.095614083 i$ \\\hline
				
				\multicolumn{2}{|c||}{Matrix method \cite{Lin:2016sch}, $\,s=\pm1$} &
				\multicolumn{2}{|c||}{Matrix method \cite{Lin:2016sch}, $\,s=\pm1$} \\\hline
				$-1$ & $0.45759551 - 0.09500443 i$ &
				$-1$ & $0.656898671 - 0.095616218 i$ \\\hline\hline
				
				\multicolumn{2}{|c||}{$\mathrm{Separate\ code},\,\varsigma=\pm1$} &
				\multicolumn{2}{c|}{$\mathrm{Separate\ code},\,\varsigma=\pm1$} \\\hline
				$+1$ & $0.37367169 - 0.088962316 i$ &
				$+1$ & $0.599443292 - 0.092703032 i$ \\\hline
				
				\multicolumn{2}{|c||}{Matrix method \cite{Lin:2016sch}, $\,s=\pm2$} &
				\multicolumn{2}{|c||}{Matrix method \cite{Lin:2016sch}, $\,s=\pm2$} \\\hline
				$+1$ & $0.37367169 - 0.088962316 i$ &
				$+1$ & $0.599443288 - 0.092703048 i$ \\\hline
			\end{tabular}
		}
	\end{table}
	
	\begin{table}[H]
		\centering
		\caption{QNMs at the Reissner-Nordstr\"om limit, where $a=0$, of our separated numerical code with $\ell=2$, $m=1$, and two distinct charges $Q=0.7$, $Q=0.99$. We compare our QNMs with those found for the same problem in Ref. \cite{Kokkotas:1988fm}.}
		\label{table:RN}
		\resizebox{\columnwidth}{!}{%
			\begin{tabular}{|c|c||c|c|}
				\hline
				\multicolumn{2}{|c||}{$Q=0.7$} &
				\multicolumn{2}{c|}{$Q=0.99$} \\\hline
				$b$ & $M\omega$ & $b$ & $M\omega$ \\\hline
				
				\multicolumn{2}{|c||}{$\mathrm{Separate\ code},\,\varsigma=\pm1$} &
				\multicolumn{2}{c|}{$\mathrm{Separate\ code},\,\varsigma=\pm1$} \\\hline
				$-1$ & $0.53651 - 0.09877i$ &
				$-1$ & $0.69275 - 0.08864i$ \\\hline
				
				\multicolumn{2}{|c||}{Method~\cite{Kokkotas:1988fm}, electromagnetic-led} &
				\multicolumn{2}{c|}{Method~\cite{Kokkotas:1988fm}, electromagnetic-led} \\\hline
				$-1$ & $0.53651 - 0.09877i$ &
				$-1$ & $0.69275 - 0.08864i$ \\\hline
				
				\multicolumn{2}{|c||}{$\mathrm{Separate\ code},\,\sig=\pm1$} &
				\multicolumn{2}{c|}{$\mathrm{Separate\ code},\,\sig=\pm1$} \\\hline
				$+1$ & $0.39250 - 0.08990i$ &
				$+1$ & $0.42930 - 0.08427i$ \\\hline
				
				\multicolumn{2}{|c||}{Method~\cite{Kokkotas:1988fm}, gravitational-led} &
				\multicolumn{2}{c|}{Method~\cite{Kokkotas:1988fm}, gravitational-led} \\\hline
				$+1$ & $0.39250 - 0.08990i$ &
				$+1$ & $0.42930 - 0.08427i$ \\\hline
		\end{tabular}}
	\end{table}
	
	\begin{table}[H]
		\centering
		\caption{QNMs at the Kerr limit, where $Q\to0$, of our separated numerical code with $\ell=m=2$, and two distinct spins $a/M=0.3$, $a/M=0.8$. We compare our QNMs with those found for the same problem in Ref. \cite{Berti:2005ys}.}
		\label{table:Kerr}
			\begin{tabular}{|c|c||c|c|}
				\hline
				\multicolumn{2}{|c||}{$a/M=0.3$} &
				\multicolumn{2}{c|}{$a/M=0.8$} \\\hline
				$b$ & $M\omega$ & $b$ & $M\omega$ \\\hline
				
				\multicolumn{2}{|c||}{$\mathrm{Separate\ code},\,\varsigma=\pm1$} &
				\multicolumn{2}{c|}{$\mathrm{Separate\ code},\,\varsigma=\pm1$} \\\hline
				$-1$ & $0.508317-0.093716i$ &
				$-1$ & $0.674180-0.079044i$ \\\hline
				
				\multicolumn{2}{|c||}{Method~\cite{Berti:2005ys}, $s=\pm1$} &
				\multicolumn{2}{c|}{Method~\cite{Berti:2005ys}, $s=\pm1$} \\\hline
				$-1$ & $0.508317-0.093716i$ &
				$-1$ & $0.674178-0.079044i$ \\\hline
				
				\multicolumn{2}{|c||}{$\mathrm{Separate\ code},\,\sig=\pm1$} &
				\multicolumn{2}{c|}{$\mathrm{Separate\ code},\,\sig=\pm1$} \\\hline
				$+1$ & $0.419527-0.087729i$ &
				$+1$ & $0.586018-0.075629i$ \\\hline
				
				\multicolumn{2}{|c||}{Method~\cite{Berti:2005ys}, $s=\pm2$} &
				\multicolumn{2}{c|}{Method~\cite{Berti:2005ys}, $s=\pm2$} \\\hline
				$+1$ & $0.419527-0.087729i$ &
				$+1$ & $0.586017-0.075630i$ \\\hline
			\end{tabular}
	\end{table}
	
	\begin{table}[ht]
		\centering
		\caption{Illustrative QNM results from the calculation that uses the decoupled equations \eqref{eq:angular-two}, \eqref{eq:radial} for KN BHs. Three cases are studied: Case (1) with $M=1$, $a=0.2$, $Q=0.4$, $m=\ell=2$, Case (2) with $M=1$, $a=0.6$, $Q=0.4$, $m=\ell=2$, and Case (3) with $M=1$, $a=0.1$, $Q=0.99$, $m=\ell=3$. All cases have varying spin label $\sig$ and branch $b$.}
		\label{table:isospectrality}
		\resizebox{\columnwidth}{!}{%
			\begin{tabular}{|c|c|c|}
				\toprule
				$\sig$ & $M\omega$ & $\lambda$\\\hline
				\multicolumn{3}{|c|}{Case (1): $b=+1$ } \\\hline
				$+1$ & $0.409918268-0.088613263i$ & $4.167148221+0.017880575i$\\
				$-1$ &  $0.409918268-0.088613263i$ & $4.167148221+0.017880575i$\\\hline\hline
				\multicolumn{3}{|c|}{Case (1): $b=-1$ } \\\hline
				$+1$ & $0.515048591-0.095571407i$ & $-4.179676202-0.012800451i$\\
				$-1$ & $0.515048591-0.095571407i$ & $-4.179676202-0.012800451i$\\\hline\hline
				\multicolumn{3}{|c|}{Case (2): $b=+1$ } \\\hline
				$+1$ & $0.5182791234-0.0825427610i$ & $3.9370634668+0.049552657i$\\
				$-1$ &  $0.5182791234-0.0825427610i$ & $3.9370634668+0.049552657i$\\\hline\hline
				\multicolumn{3}{|c|}{Case (2): $b=-1$ } \\\hline
				$+1$ & $0.6267598372-0.0878857714i$ & $-4.0059969837-0.0315888614i$\\
				$-1$ & $0.6267598372-0.0878857714i$ & $-4.0059969837-0.0315888614i$\\\hline\hline
				\multicolumn{3}{|c|}{Case (3): $b=+1$ } \\\hline
				$+1$ & $0.7529019146-0.07928136606i$ & $-3.984883079+0.7888525253i$\\
				$-1$ &  $0.7529019146-0.07928136606i$ & $-3.984883079+0.7888525253i$\\\hline\hline
				\multicolumn{3}{|c|}{Case (3): $b=-1$ } \\\hline
				$+1$ & $1.00961419653-0.0828574161i$ & $-7.018566494+0.7721881108i$\\
				$-1$ & $1.00961419653-0.0828574161i$ & $-7.018566494+0.7721881108i$\\\hline\hline
		\end{tabular}}
	\end{table}
	
	The label $b$ specifies the initial angular branch and need not coincide with the sign of $\operatorname{Re}\lambda$ away from the reference limit.
	
	\bibliography{biblio}
	
\end{document}